\documentclass{elsart}
\usepackage{graphicx}
\usepackage{amssymb}
\begin{document}
\begin{frontmatter}
\title{The Circulation of Money and Holding Time Distribution
}
\author{Yougui Wang\corauthref{cor1}
}
\author{, Ning Ding,}
\author{Li Zhang}
%

\corauth[cor1]{Corresponding author.
\\ {Tel.: +86-10-62207876; fax:+86-10-62207876.}
\\{\em E-mail address:\/}\, ygwang@bnu.edu.cn}
\address{Department of Systems Science, School of Management, Beijing Normal University, Beijing,
100875, People's Republic of China
}

\begin{abstract}
We have studied the statistical mechanics of money circulation in
a closed economic system. An explicit statistical formulation of
the circulation velocity of money is presented for the first time
by introducing the concept of holding time of money. The result
indicates that the velocity is governed by behavior patterns of
economic agents. Computer simulations have been carried out in
order to demonstrate the shape of the holding time distribution.
We find that, money circulation is a Poisson process in which the
holding time probability distribution follows a type of Gamma
distribution, and the velocity of money depends on the share for
exchange and the number of agents.

\end{abstract}


\end{frontmatter}
\section{Introduction}
Since there are so many analogies between some features of economy
and objects of physics, some advanced physical methods could be
applied into analyzing economic issues \cite{1}.In particular, the
application of statistical physics methods to financial market has
achieved a great success \cite{2}. Recently, many efforts are
being made to expand the scope of such analogical applications to
other economic problems \cite{3,4}. Among these problems, the
dynamics of money has attracted much attention because of its
significance in economy \cite{5,6,7}. By identifying money in a
closed economy with energy and the average amount of money with
temperature, Adrian Dr\"{a}gulescu and Victor Yakovenko have
demonstrated that the equilibrium probability distribution of
money follows the exponential Boltzmann-Gibbs law due to the
conservation of money \cite{6}. The distribution of money in a
real economy mainly depends on economic agents' behaviors and the
manner they interact between each other. According to Adrian
Dr\"{a}gulescu and Victor Yakovenko, if a model has time-reversal
symmetry, then a stationary Boltzmann-Gibbs distribution can be
obtained; otherwise, the system may have a non- Boltzmann-Gibbs
distribution or no stationary distribution at all. Following this
work, an example of stationary non- Boltzmann-Gibbs distribution
is given by Anirban Chakraborti and Bikas K. Chakrabarti when the
impact of the saving propensity of the agents on the statistical
mechanics of money is taken into account \cite{7}.

Indeed, money does matter in the performance of economy. The
statistical analysis of money is therefore essential and very
intriguing. However, comparing with its distribution among
economic agents, understanding of the motion of money in the
economy is more significant. As medium of exchange and unit of
accounting, money cannot be used up as consumption goods or worn
off as capital goods, but can only be transferred between agents.
So it can be assumed to be conserved after being injected into an
economy. The picture that money moves from hand to hand between
agents in sequence draws some physicists to use one-dimensional
lattice or square lattice network models to deal with monetary
issues \cite{4,5}. Although this analogy might appear attractive
intuitively, it is not a short-cut for addressing these issues. In
fact, economists have been thinking about the money in an
alternative way by placing themselves in an economic space instead
of usual physical one. The economic space comprises only two
parts: demand and supply. Correspondingly, the economic agents can
be classified to two groups: buyers and sellers. The former who
holds money is in the demand side and the latter who produces
goods is in the supply side. In economic space, the motion of
money then can be figured out as moving between these two regions.
When exchange occurs in the market, money flows from demand region
to supply region can be seen. After that, money in the supply
region is taken back to the demand region immediately, because the
seller who just participates in the exchange becomes a potential
buyer. It's so-called circulation of money that the process takes
place continuously.

The most representative economic theory that copes with money in
such a way is the quantity theory of money \cite{8}, which
describes the relationship between money flow and product flow at
macro level. As the total amount of money is conserved, the money
flow originates from monetary circulation. Meanwhile money is
circulating; goods are continuously produced, exchanged and
consumed, in which the product flow arises. Thus it can be seen
that both flows coexist all the time and the ratio of them is
called price level. Given the price level fixed, the money flow
corresponds to the product flow. So the circulation of money can
reflect the operation of production economy. Furthermore, in a
market economy effective demand is always deficient and the real
output is dominated by willing expending. In this case, the
circulation of money is not only the reflection but also the
driving force of production. This is the reason that the quantity
theory of money has been of high reputation in economics since
being set forth.

In the quantity theory of money, in order to express the
relationship between money and production, a significant variable,
the velocity of money circulation, has been introduced. Although
its definition is explicit, the essence of this concept remains
ambiguous. The velocity of money circulation is defined as the
ratio of money flow to the money stock. It can be calculated with
nominal gross domestic product (GDP) and total amount of money.
There is however no consensus so far on what factors govern this
variable. Our goal is to describe the velocity of money by
utilizing the ideas and concepts from statistical physics and to
show the statistical characteristics of money circulation by
computer simulations.

The statistical description of the velocity of money circulation
is based on holding time of money which is defined as time
interval between two transactions. Although this concept is kept
in mind when economists think of the velocity, even the term
referring to this kind of time interval has been mentioned in
several cases, it's somewhat new to them since there has been no
explicit specification of it in economics. While there has been a
similar term in physics which is called waiting-time. In recent
times, several efforts have been devoted to measure the waiting
time distributions in financial markets, see e.g. \cite{9,10}. In
the process of money circulation, not only the amount of money
each agent holds can be considered as random variables, but also
the holding time between two transactions varies randomly. Thus
there exists a holding time distribution of money in a closed
system and the connection between holding time and the velocity is
established. The holding time is not a physical time but an
economic variable which is determined by economic agents'
behavior. Consequently, with the introduction of holding time the
velocity can be expressed in a statistical way, and furthermore a
bridge that links such a macroeconomic variable to individual
agents' choices might be set up.

We find that the velocity of money is the expectation value of the
reciprocal of holding time. Thus the characteristics of the
holding time distribution do matter. Due to absence of the data of
money circulation, empirical study is not feasible. Therefore, we
performed computer simulations to see the shape of the holding
time probability distribution and demonstrated how the individual
choices and other factors affect the velocity.

Our model is a simple extension of Adrian Dr\"{a}gulescu and
Victor Yakovenko's work, in which random exchange approach
remains. However, we placed emphasis on measuring holding time
distribution instead of money distribution. The dependences of the
velocity of money circulation on some governing factors were
discussed theoretically and experimentally.

\section{The Quantity Theory of Money}
When physicists intend to study economic issues, it is helpful to
see how economists deal with them. Having knowledge about relevant
basic thoughts and their context that economists have already
developed enables us to apply physics to those issues more
effectively. The circulation of money has been talked about for
hundreds of years; however, it is still a topic of great interest
with some puzzles uncovered. Many publications about it can be
found in economics literature. In this section we shall present a
brief review of the basic knowledge and the main results of the
theory.

The quantity theory of money is a well-known doctrine to
economists which emphasizes that the money supply is the main
determinant of nominal GDP. This theory is constituted of two
branches: one is built on the Fisher exchange equation; the other
is on the Cambridge cash balance equation. The Fisher equation
places emphasis on the part as medium of transactions that money
plays and states that \cite{11}
\begin{equation}\label{fisher}
MV=PY,
\end{equation}
where $M$ is the amount of money in circulation, $V$ the velocity
of circulation of money, $P$ is the average price level and $Y$ is
the level of real income. The left hand side of the equation
represents the amount spent on final goods and services in
transactions while the right hand side represents the amount
received for these goods and services. In other words, the left
hand side is the total spending from a monetary perspective while
the right hand side is that from a real view. By definition, these
two sides must be equal. Thus this equation shows the relationship
between money flow and product flow. The Cambridge equation
proposes that money is desired as a store of value and regarded
money demand as a function of nominal income, i.e. \cite{12}
\begin{equation}\label{cambridge}
M=kPY,
\end{equation}
where $k$ is the famous ``Cambridge constant".

From Eq.(\ref{fisher}) we can immediately obtain
\begin{equation}\label{Delta}
\frac{\Delta M}{M}+\frac{\Delta V}{V}=\frac{\Delta
P}{P}+\frac{\Delta Y}{Y}.
\end{equation}
But it is only a matter of arithmetic, not of economics. Till now
we have not made any assumption of the causality between left and
right.

According to Fisher, the statement of the quantity theory requires
three pillars: firstly, that $V$ and $Y$ are fixed with respect to
the money supply. Secondly, that the supply of money is exogenous.
Thirdly, the direction of causation runs from left ($MV$) to right
($PY$). The story of the quantity theory then runs like this:
since $V$ and $Y$ are fixed and $M$ is exogenous, then an increase
in the supply of money will lead to an exactly proportionate
increase in the price level. Thus, money supply expansions only
cause price inflation. This is the so-called monetary neutrality.
For reaching this, the last pillar is the most crucial one among
the three mentioned above.

Comparing equations (\ref{fisher}) and (\ref{cambridge}), if we
simply set $k=1/V$, there seems be no difference between these
two. In fact, the Cambridge story, however, is fundamentally
different from the Fisher story. The proposition that the
Cambridge advances comes from an implied hypothesis that the
direction of causation runs from right ($kPY$) to left ($M$). The
reverse of causation leads to a possible unstable ``Cambridge
constant", that is to say, change in money supply may have real
effects.

Following this line of the Cambridge approach, monetarists present
a restatement of the quantity theory of money \cite{13}. But their
fundamental contributions to the development of the theory are the
empirical researches on this aspect \cite{14}. Basing on this
work, they believe that the velocity of money is stable.

Nowadays, as to the causality of money supply and the real output,
most economists believe that the monetary neutrality holds in the
long run (a few years). As Milton Friedman once made the
statement, ``Inflation is always and everywhere a monetary
phenomenon." In the short run, however, they are not in agreement
about whether changes in the money supply lead to changes in the
price level or in the real output.

What on earth is the causality of money supply and the real
output? And why the monetary neutrality is correct in the long run
but not in the short run? The finish of solving these questions
lies on the thorough understanding to the nature of the
circulation of money. In fact, with the development of the
quantity theory of money, people understand the velocity step by
step; in other words, the process of developing the quantity
theory of money is actually that of comprehending the velocity.
The earliest version of the quantity theory of money takes the
velocity as an exogenous variable and being determined by
institutional arrangements and technologies of transaction. As a
result, it can be presumed as a constant. Cambridge economists
think that money is a kind of asset and the holders of them tend
to optimize their portfolio. So the factors related to money
demand are considered when investigating the velocity. Basing on
that, Friedman dedicated precisely to unearthing the relationship
between the velocity and those factors such as interest rate,
expected inflation, permanent income, and the return on money,
etc. However, all of analysis mentioned above is only qualitative,
and no concrete function for expressing that relationship is
offered. In the following section, we shall present a statistical
expression of the velocity of money.

\section{Statistical Analysis of The Circulation Velocity of Money}
\subsection{Holding time of money}
The advantage of money as the medium of exchange, in that it
overcomes the need to obtain coincidence of wants; it implies that
an agent can sell his good at one time for ``money" and then
trading his ``money" for the goods he finally wishes to purchase
at another time. The divorce of sale and purchase results in a
time interval within which money stays in the pocket of agents.
Moreover, money is an asset which allows value to be stored
easily, and there is also a time interval for agents to preserve
it. In the real economy, money is actually changing hands all the
time. In this process, if an agent receives money from others at
one moment, he will hold it for a period, and eventually pays
something to another agent. Now we introduce a new concept named
as holding time which is defined the interval between the receipt
of income and its disbursement.

Holding time is not an intrinsic character of money itself, but is
a character of its holder's behavior in utilizing the money for
certain purpose. The reasons about why people hold money arise
from the composite result of a number of different motives.
Economists provide several variations of explanations about it.
According to traditional approach, the motives are classified
under three headings: the transactions-motive, the
precautionary-motive and the speculative-motive. Keynes classified
them under four headings by further dividing the
transactions-motive into the income-motive and the business-motive
\cite{15}. Despite what motives are taken into account and how
they are classified, the existence of holding time could not be
denied. Their differences are only embodied in determinants of the
holding time.

In the circulation story, money circulates round and round in
economic space. At the beginning of one cycle, money stays in the
demand region till it participates in the transaction. When the
exchange takes place, it moves to the supply side and immediately
goes back to the initial position and gets to the end of this
cycle or the beginning of the next. Since it is reasonable to
assume that the transaction does not take any time, the holding
time is equal to the period of one cycle that money goes.

\subsection{Distribution of money over holding time}
Let's consider the economy in which the total quantity of money is
$M$. At any given time, each unit of them must have ever
participated in the exchange at certain previous moment and will
repeat at a certain moment in the future. Thus it has its own
specific holding time at present. As a result, the money may have
different holding time at the same time, due to either different
holders or various motives of the same holder. Therefore, the
conserved money spreads over the holding times.

Now we introduce the probability distribution function of money
$P(\tau)$, which is defined the amount of money whose holding time
is between $\tau$ and $\tau+\,d\tau$ is equal to
$MP(\tau)\,d\tau$. For any unit of money, $P(\tau)$ is actually
the probability of taking part in exchange after an interval of
$\tau$. Thus we can give the normalization condition as follows:
\begin{equation}\label{normalization}
\int\nolimits^\infty_0 P(\tau)\,d\tau=1.
\end{equation}
And we can also have the expectation of the holding time of money,
$\bar\tau$.
\begin{equation}\label{averagetau}
\bar\tau=\int\nolimits^\infty_0 P(\tau)\,\tau\,d\tau.
\end{equation}
We call it the average holding time of money.

Each unit of money injected into the economy changes hands for
countless times. The time interval between any two times is
various because the holders may be different. That is to say, the
holding time of the same unit may be altered at different cycle
due to the change of holder. It follows that the distribution of
money over the holding time evolves with time. If the behavior
pattern of the economic agents keeps unchanged, we can get an
equilibrium state where the distribution of money is stationary.
In other words, any single money's holding time may strongly
fluctuate, but the overall probability distribution does not
change.

In the stationary state, the fraction of money $MP(\tau)\,d\tau$
participates in the exchange after a period of $\tau$. The money
flow generated by this fraction then is
\begin{equation}\label{tauflow}
F(\tau)=MP(\tau)\,\frac{1}{\tau}\,d\tau.
\end{equation}
The total money flow $F$ in the economy is equal to the sum of the
money flows due to the contribution of all parts whose holding
time is equal to $\tau\,(\tau\in[0,\infty)\,)$, then we have
\begin{equation}\label{moneyflow}
F=M\int\nolimits^\infty_0 P(\tau)\,\frac{1}{\tau}\,d\tau.
\end{equation}
The velocity indicates the speed at which money circulates. From
its definition $V=F/M$, we get
\begin{equation}\label{velocity}
V=\int\nolimits^\infty_0 P(\tau)\,\frac{1}{\tau}\,d\tau.
\end{equation}
This is the statistical expression of the circulation velocity of
money with respect to holding time. The result shows the velocity
$V$ is the mathematical expectation of $1/\tau$. From
Eqs.(\ref{averagetau}) and (\ref{velocity}), it is obvious that
the velocity and the average holding time are reversely related.
It also indicates that all the impacts of factors related on the
velocity take effect through influencing the economic agents'
choices of holding time. Thus, an effective measure for increasing
(decreasing) the velocity of money is to motivate the economic
agents to shorten (lengthen) the average holding time.

The result that $P(m)$ follows Boltzmann-Gibbs' distribution arise
from non-negativity of the money amount possessed by each unit of
agent and conservation of the total money. But till now we can not
say anything about what the distribution $P(\tau)$ of money is
likely to be. In order to see the characteristics of $P(\tau)$, we
performed several computer simulations which will be described in
next section.

\section{Model and Computer Simulations}
The model we used is very similar to that of Adrian Dr\"{a}gulescu
and Victor Yakovenko. The economic system is closed where the
total money $M$ and the number of economic agents $N$ are fixed.
Each of agents has a certain amount of money initially. The money
exchange is performed by agents in sequence. In every round an
arbitrary pair of agents is chosen to exchange, and the amount of
money that changes hands is given according to the trading rule.
The non-negativity of any agent's money and the conservation of
the total money in each round are ensured.

We carried out the measurement by employing the trading rule which
has been used by Adrian Dr\"{a}gulescu and Victor Yakovenko in
their simulations. Initially, the total money $M$ is divided
amongst $N$ agents equally so that each agent possesses the same
amount of money $M/N$. We choose a pair of agents randomly at a
time; one of them is randomly picked to be the ``receiver" , then
the other one becomes the ``payer", and the amount $\Delta m$ is
transferred from the payer to the receiver. The trading rule lets
the exchange amount in each round be of the following form:
$\Delta m=\frac{1}{2}\nu(m_1+m_2)$, where $\nu (0<\nu<1)$ is a
random fraction, $m_1$ and $m_2$ are the amount of money possessed
by the payer and the receiver, respectively. If the payer can't
offer the amount, the transfer dose not take place, and we turn to
another pair of agents.

Following this procedure, the final stationary distribution of
money among agents $P(m)$ is Boltzmann-Gibbs' distribution, which
is shown in Fig. 1. Whether the system is at stationary state or
not is judged by observing the evolution of the entropy
$S=-\int\nolimits^\infty_0\,dmP(m)\ln P(m)$. This is illustrated
in the inset of Fig. 1. All the records are obtained after $S$
reaches the maximal value denoted as the vertical line in the
inset of Fig. 1, and average over 400 such stationary
distributions was taken to obtain a smooth distribution.

The main purpose of our simulation is to show the holding time
distribution of money $P(\tau)$ instead of $P(m)$. In order to
obtain the holding times of all money, we numbered all of the
money and tracked each of them. The moment at which each unit of
money participated exchange for the latest time was memorized.
After majority (99\%) of money had ever taken part in trade, we
began to measure holding times by recording the corresponding
moment at which each unit of money participated exchange for the
first time hereafter. The holding time of each unit is the
difference between the two moments. Due to the random exchange,
there are always a few units whose holding time is too long to be
recorded. Thus we just sampled the majority of total money, and
the remainder was omitted. After that we got the distribution of
money over holding time.

\section{Results and Discussion}
We performed several simulations of $P(\tau)$ by altering values
of $N$ and $M$. The results show that the holding time
distributions of money we observed are stationary and have
remarkably similar form independent of the values given. The
typical one is illustrated in Fig. 2 where the size was taken to
be $N=2500$ and $M=250000$. However, the profile of this
distribution cannot provide any clue to make sure which kind of
distribution it belongs to.

To examine the origination of this distribution, we turned to
observe another kind of temporal distribution of money. Taking the
time when we began to record as zero point, hereafter, we set the
moment at which the unit of money is transferred for the first
time to be latency time of this unit. Please notice that the
latency time we defined here is different from the holding time of
money. Fig.3 shows the latency time distribution of money for this
case. It can be seen from the inset of Fig. 3 that the
distribution follows exponential law:
$P(t)=\frac{1}{T}e^{-\frac{t}{T}}$. The result we obtained here
indicates that the transferring process of money is a Poisson
process with the rate of changes $1/T$. Thus the average latency
time is equal to $T$ and its value obtained in Fig. 3 is about
5550.

Since the whole process is an independent stationary stochastic
one, when we begin to observe one unit of money, if we look
forward from the zero point of time, the probability of that it
takes part in exchange during the period between $t$ and $t+dt$
for the first time is $P(t)\,dt$; if we look backward, the
probability of that the unit of money has ever been traded during
the period between $-t$ and $-t+dt$ and has been held till $t=0$
is also $P(t)\,dt$. Suppose the unit of money takes part in
exchange at $t_1 (t_1>0)$ and its holding time is $\tau (\tau\geq
t_1)$, then it must be transferred at moment $t_1-\tau$ . As a
result, the probability of this case is $P(t_1)P(\tau-t_1)\,dt_1$,
and the probability of that the holding time is equal to $\tau$
can be obtained
\begin{equation}\label{poftau}
P(\tau)=\int\nolimits^\tau_0P(t_1)\,P(\tau-t_1)\,dt_1=\frac{1}{T^2}\,\tau
e^{-\frac{\tau}{T}}.
\end{equation}
This result shows the probability distribution of money over
holding time follows a type of Gamma distribution.

Substituting Eq.(\ref{poftau}) into (\ref{velocity}), we get
\[
V=\frac{1}{T}.
\]
It's not surprising to this result for the velocity of money is
just the rate parameter of the circulation process. We further
more substitute Eq.(\ref{poftau}) into (\ref{averagetau}), then
\[
\bar\tau=2T.
\]
Thus the relationship between the velocity and the average holding
time can be simply expressed as follows:
\[
\bar\tau=\frac{2}{V}.
\]
We proceeded a fitting of the simulation result in Fig. 2 with the
theoretical expression of Eq. (\ref{poftau}) and found a good
agreement. The fitting value of the average latency time is about
5548, which is almost the same as the result derived from Fig. 3.
The deviation may be due to nonuniform distribution of random
number generated in the process of simulation.

To consider how individual agent's choice affects the velocity of
money, we furthermore added a multiplier $k$ and denominated it
share for exchange, the exchange amount then becomes $\Delta
m=\frac{k}{2}\nu(m_1+m_2)$. As a further check we have determined
the dependencies of the velocity of money on the share for
exchange $k$, the number of agents $N$ and the average amount of
money in the system $\bar m$, which are shown in Figs. 4(a)-(c)
respectively. For each case, we performed the simulation and got a
holding time distribution of money $P(\tau)$, after that the
corresponding velocity was simply deduced by using
Eq.(\ref{velocity}).

In Fig. 4(a) we set $N=2500$ and $\bar{m}=100$. It can be seen
that the velocity is proportional to the share $k$. For larger
share for exchange the velocity of money in the trade process
increases. So we can conclude that the circulation velocity is
determined by the agents' behaviors in the exchange since the
share for exchange $k$ reflects economic agent's choice. Fig. 4(b)
shows the velocity of circulation $V$ plotted versus $1/N$, for
$k=1$ and $\bar{m}=100$. As shown, $V$ vs $1/N$ is linear for the
whole range of $N$. Fig. 4(c) shows the variation of the velocity
of circulation with average amount of money $\bar{m}$ for $k=1$
and $N=2500$, from which we can see that $V$ remains constant for
different values of $\bar m$.

These results can be understood by a simple deduction. According
to the trading rule: $\Delta m=\frac{k}{2}\nu(m_1+m_2)$ and
non-negativity of $m$, the average money flow, $F$, can be derived
as follows:
\[
F\propto k\bar m.
\]
From its definition it can also be written as
\[
F=MV=N\bar{m}V.
\]
Comparing these two equations above, we immediately get
\[
V\propto \frac{k}{N}.
\]
Thus the velocity of money is proportional to the share for
exchange, reversely proportional to number of agents, and
independent of average amount of money. This theoretical result is
in good agreement with that of simulations.

\section{Summary}
In this paper, we have considered money circulation in a closed
economy by applying the statistical approach. We present the
quantity theory of money briefly to argue that the circulation
velocity of money is an essential variable for our understanding
of the dynamics of money. By introducing the concept of holding
time of money, we provide a statistical expression of the
circulation velocity. The result indicates that the velocity and
the average holding time are reversely related and the main
determinant of them is the agents' behaviors.

We have performed several computer simulations based on random
exchange model. We find the money involved in exchange process
possesses not only a stationary probability distribution among
agents but also a stationary one over holding time. The holding
time probability distribution is found to be a type of Gamma
distribution because the circulation of money is a kind of Poisson
process. The dependence of the circulation velocity of money on
agents' choices are demonstrated by changing the share for
exchange. We also investigated the dependence of the distribution
on the number of agents and the average amount of money per agent.
The theoretical results we derived according to the model show
good agreements with the simulations. We believe that this study
promises a fresh insight into the velocity of money circulation
and opens a way to a firm microfoundation of it.

\section*{Acknowledgments}
This research was supported in part by the National Science
Foundation of China under Grant No. 70071037. The authors thank
Zengru Di for stimulating discussions and suggestions.

\newpage
\section*{Figure Captions}
\begin{description}

\item[Figure 1] The stationary probability distribution of money
among agents $P(m)$ versus money $m$. Solid curve: fit to the
Boltzmann-Gibbs law $P(m)\propto exp(-m/\bar{m})$. The vertical
line indicates the initial distribution of money. The inset shows
time evolution of entropy S and the vertical line denotes the
moment after which the measurements are performed.
\\
\item[Figure 2] The stationary probability distribution of money
versus holding time $P(\tau)$. The solid curve is fit to the
equation $P(\tau)=\frac{1}{T^2}\tau exp(-\tau/T)$ with $T=5548$.
\\
\item[Figure 3]The stationary distributions versus latency time
$t$. The fitting in the inset indicates that the distribution
follows the exponential law: $P(t)=\frac{1}{T}exp(-t/T)$.
\\
\item[Figure 4]  Dependencies of the velocity of money circulation
$V$ on (a) the share for exchange $k$; (b) the reciprocal of
number of agents $1/N$; (c) the average amount of money in the
system $\bar{m}$.
\end{description}
\clearpage
\begin{figure}
\includegraphics[width=\textwidth]{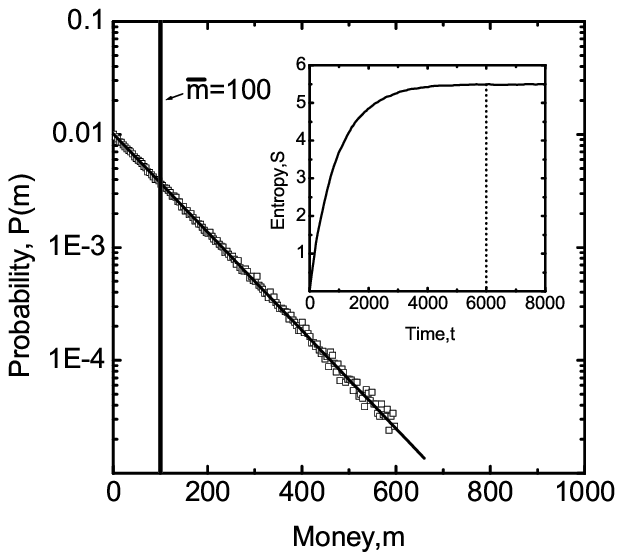}
\centering{\Large{Wang and Ding, Fig. 1}}\label{Fig1}
\end{figure}
\clearpage
\begin{figure}
\includegraphics[width=\textwidth]{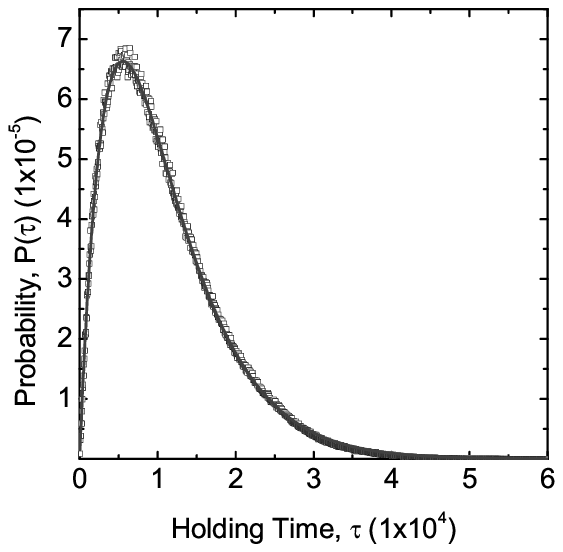}
\centering{\Large{Wang and Ding, Fig. 2}}\label{Fig2}
\end{figure}
\clearpage
\begin{figure}
\includegraphics[width=\textwidth]{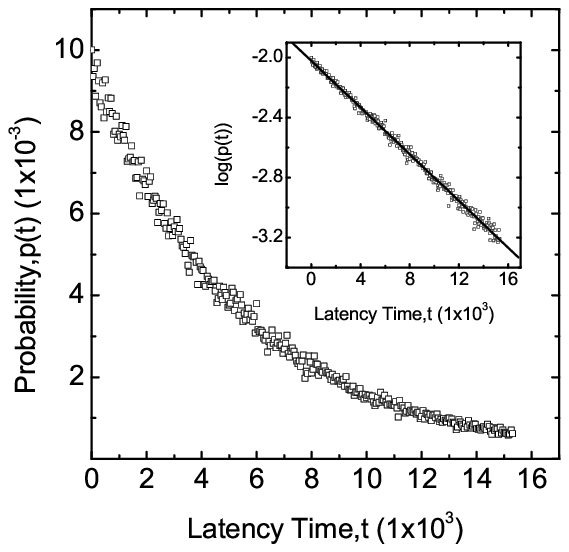}
\centering{\Large{Wang and Ding, Fig. 3}}\label{Fig3}
\end{figure}
\clearpage
\begin{figure}
\includegraphics[width=\textwidth]{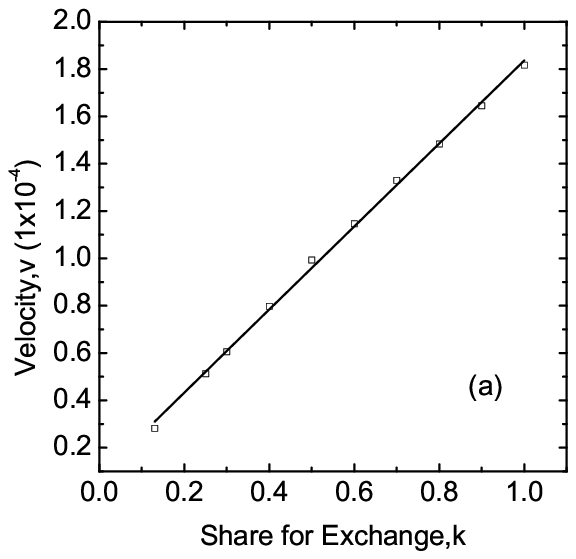}
\centering{\Large{Wang and Ding, Fig. 4(a)}}\label{Fig4a}
\end{figure}
\clearpage
\begin{figure}
\includegraphics[width=\textwidth]{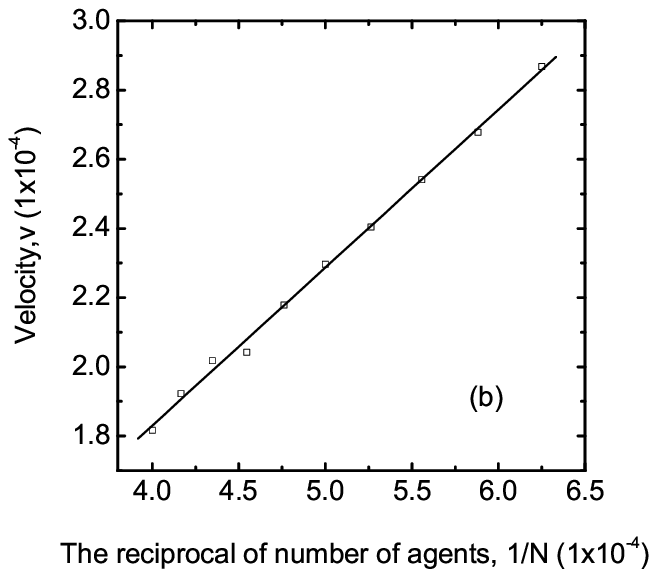}
\centering{\Large{Wang and Ding, Fig. 4(b)}}\label{Fig4b}
\end{figure}
\clearpage
\begin{figure}
\includegraphics[width=\textwidth]{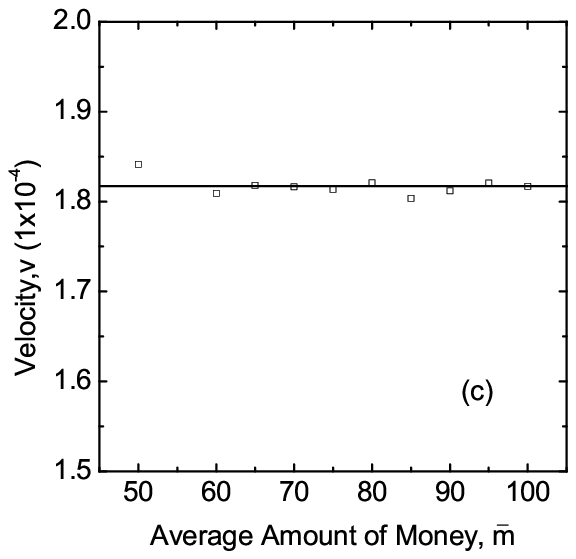}
\centering{\Large{Wang and Ding, Fig. 4(c)}}\label{Fig4c}
\end{figure}
\end{document}